\documentclass[prb,aps,showpacs,twocolumn,amsmath,amssymb]{revtex4}

\usepackage{graphicx}
\usepackage{amssymb,amsmath}

\begin{document}

\title{Graduate quantum mechanics reform}

\author{L. D. Carr}
\affiliation{Department of Physics, Colorado School of Mines, Golden, Colorado, 80401}

\author{S. B. McKagan}
\affiliation{JILA, University of Colorado and NIST, Boulder, Colorado 80309, USA}


\begin{abstract}
We address four main areas in which graduate quantum mechanics education can be improved: course content, textbook, teaching methods, and assessment tools. We report on a three year longitudinal study at the Colorado School of Mines using innovations in all these areas. In particular, we have modified the content of the course to reflect progress in the field in the last 50 years, used textbooks that include such content, incorporated a variety of teaching techniques based on physics education research, and used a variety of assessment tools to study the effectiveness of these reforms. We present a new assessment tool, the Graduate Quantum Mechanics Conceptual Survey, and further testing of a previously developed assessment tool, the Quantum Mechanics Conceptual Survey. We find that graduate students respond well to research-based techniques that have been tested mainly in introductory courses, and that they learn much of the new content introduced in each version of the course. We also find that students' ability to answer conceptual questions about graduate quantum mechanics is highly correlated with their ability to solve calculational problems on the same topics. In contrast, we find that students' understanding of basic undergraduate quantum mechanics concepts at the modern physics level is not improved by instruction at the graduate level.
\end{abstract}


\maketitle

\section{Introduction}
\label{sec:introduction}

There are significant problems with graduate quantum mechanics education in the U.S., both in content and pedagogy. The current content is outdated and inadequate for preparing graduate students to be researchers in any field based on quantum mechanics, and the standard pedagogy is based on a model that research has shown to be ineffective for student learning.\cite{Hake1998a,Hrepic2007a} This article investigates a remedy for these difficulties in the context of a specific new course model.

The history of quantum mechanics can be divided into four main periods according to its major developments and applications. In period I, comprised of roughly ten years after the formulation of the Schr\"odinger equation in 1926,\cite{schrodingerE1926,Schrodinger1926a} Heisenberg, Schr\"odinger, Born, Dirac, von Neumann, and others worked out most of the material we now present in undergraduate courses, including wave mechanics, the matrix formulation, and elementary dynamics.\cite{sakurai1994,Landau1977a} By the end of this time the question of the interpretation of quantum mechanics was dropped by most researchers.\cite{EverettH1957,wheelerJA1957} In period II quantum mechanics was applied in a number of new contexts, from fields to many-body theory, leading to the development of quantum electrodynamics\cite{feynman1998} and BCS theory\cite{bardeen1957} among other triumphs. Period~III begins in 1964 with Bell's theorem.\cite{bellJS1964} The question of interpretation was renewed, and many theoreticians devised new understandings of quantum mechanics.\cite{bohmD1995} Still, experimental methods to address such questions were mostly lacking. Then, in 1982 Aspect made the first conclusive, widely discussed measurement of Bell's inequalities.\cite{aspect1981,aspect1982} This launched period IV of quantum mechanics, which includes quantum information processing and entanglement as a fundamental concept of quantum theory.

Let us be specific as to the difficulties with present graduate quantum mechanics education in the U.S. Consider, for instance, the text by Sakurai\cite{sakurai1994} which is often used for a year-long course in the first year of graduate school, treats periods I and II of quantum mechanics. Even if one can argue that the theoretical considerations of period III are too abstruse for a first year graduate course, the experimental developments of period IV belie this line of reasoning. We are convinced, based on our personal experience, that it is not possible to follow new developments in and contribute to any field in which quantum theory is a fundamental ingredient without a knowledge of all four periods of quantum mechanics. From experimental techniques in quantum computing and other applications of quantum mechanics to theoretical advances such as density matrix renormalization group methods, an understanding of periods III and IV is essential. Yet texts such as Sakurai leave out any mention of periods~III and IV. At best, these subjects are treated as an afterthought in many texts. These afterthoughts are typically omitted or only briefly mentioned in courses.

Is it possible to include a more complete picture of quantum mechanics in a first year course? For example, can one provide a significant treatment of entanglement and the density matrix formalism at the first year graduate level? One might argue that such material belongs in a second year course on quantum field theory or advanced quantum mechanics. However, in our experience, and as reflected in textbooks such as Peskin and Schroeder,\cite{peskinME1995} such courses restrict themselves to period II with a few later developments in the spirit of the first two periods of quantum mechanics.

To address the inadequacies in both content and pedagogy in standard graduate quantum mechanics instruction, we have incorporated numerous reforms into a graduate quantum mechanics course and have used a variety of assessment tools to examine the effect of these reforms. Our study addresses the following questions: (1) How can we incorporate up-to-date material in an actual course -- what needs to be cut and what added? (2) What graduate-level quantum mechanics textbooks, if any, incorporate periods I through IV of quantum mechanics in a pedagogically sound and readable way? (3) What teaching innovations inspired by physics education research (PER) at the undergraduate level can be incorporated into a graduate course? (4) How can we obtain better information about what students actually learn and how it affects their attitude, without over-burdening the instructor? Finally, we have as an overarching directive for our study, what will best serve the research needs of graduate students in our specific programs? We make this last point because we believe it is paramount to keep in mind this primary motivation for doctoral-level graduate education.

Our strategy for reform is as follows. We performed a longitudinal study over three years. In the first year we taught a standard benchmark course in graduate quantum mechanics based on Sakurai. In the the following two years we taught two different courses using new textbooks that incorporate all four periods of quantum mechanics. Each of these textbooks had very different educational strategies, as discussed in Sec.~\ref{sec:content}. Most of the lecture notes were written from scratch in years two and three. Throughout the three years we gradually increased use of PER-based teaching techniques, as discussed in Sec.~\ref{sec:methods}. The same assessment tools were used all three years. As we will discuss in Sec.~\ref{sec:assessment}, we used seven different assessments, some established and some novel. In Sec.~\ref{sec:conclusions} we present the conclusions of our study and the outlook for future such investigations.

This study was performed at the Colorado School of Mines (CSM), a small, highly competitive, technically-oriented university. About a quarter of the graduate population of the physics department at CSM consists of non-traditional students who did not follow the standard track from high school diploma to undergraduate degree to graduate school, and/or have families with children, a category in which CSM can excel because many excellent non-traditional students are overlooked by larger departments. During all three years the graduate quantum mechanics course included about 10 doctoral students who took the entire year, and about 10 masters students, along with the occasional ambitious undergraduate, most of whom took only the first semester. Although the first semester is required for both M.S.\ and Ph.D.\ students, only the PhD students are required to take the second semester. The course was taught by an assistant professor (LDC) whose main area of research is quantum many-body theory, and who was teaching graduate quantum mechanics for the first time.

\section{Course Content}
\label{sec:content}

The reform of course content is divided into the modified syllabus and the associated textbooks to support the syllabus. Throughout this section and Sec.~\ref{sec:methods} we will reference the seven assessment tools of Sec.~\ref{sec:assessment} to back up our statements. In brief these are the Quantum Mechanics Conceptual Survey (QMCS), the Graduate Quantum Mechanics Conceptual Survey (GQMCS), university-wide evaluations, written student evaluations, student interviews during the course, faculty evaluation by the Department Head, and evaluation of the faculty by physics education researchers.

\subsection{Syllabus}
\label{ssec:syllabus}

In year one, which served as the benchmark course against which to compare the reforms made in years two and three, we followed a standard syllabus based on Sakurai that covered only the first two periods of quantum mechanics. In years two and three of our longitudinal study we added the following material to help students succeed in research: a clear and careful statement of the postulates of quantum mechanics; a thorough treatment of the density matrix formalism, partial traces and entanglement; introduction to quantum field theory; quantum fluctuations and the Casimir-Polder effect; the Wigner theorem and more advanced symmetry treatments;
and a wider breadth of applications in nuclear physics, optics, and solid state physics. In the second year we also added a discussion of polarization states and the classical limit for bosons, and a more
formal treatment of Hilbert spaces in infinite dimensions, including unbounded operators. These last two additions were unpopular, as evidenced by student complaints in interviews and student written evaluations. They were also confusing, as evidenced by in-class discussions and student answers to oral questions posed by the instructor. Thus, we dropped them in the third year. The particular areas of application, namely, nuclear physics, optics, and solid state physics, were motivated by the emphasized areas of research in the CSM physics department. Therefore, they satisfied our goal of directing all course reform toward helping graduate students do better in their doctoral work. These application areas could be adjusted depending on where the reformed course is being taught.

To accommodate these additions it was necessary to omit some standard material from the course. We cut the following topics, which were included in year one, from the second and third years: the full-blown treatment of the Wigner-Eckart theorem and irreducible tensor operators; inelastic scattering; a review of undergraduate quantum mechanics; some atomic, molecular, and optical applications; the calculation of arbitrary Clebsch-Gordan coefficients and Young tableaux; some scattering examples including the eikonal approximation and the low energy $s$-wave limit; and all reviews of mathematical physics, such as classical rotations.

One course goal, stated explicitly in the syllabus, was understanding three model systems: the two-state system; the harmonic oscillator; and the free particle. These systems were used as the keys to understanding dynamics and resonance, elementary quantum field theory, and scattering. There was a strong emphasis on the Dyson equation and the path integral formalism as significant mathematical techniques. Because a mathematical physics course is required as part of the graduate program in the first semester, it was unnecessary to teach basic mathematical skills such as coordinate system representations and contour integration. Further details are specific to the choice of textbook, as discussed in Sec.~\ref{ssec:textbook}.

The complete final syllabus for years two and three of our study is presented in Appendix~A.

In both years the texts enabled easy implementation of the major changes in the course content. Students enjoyed all three courses, as stated in written student evaluations. From a performance standpoint they did better on specific questions in the GQMCS, for example on questions 1, 2, 7, 9, and 10 (see Appendix~A) in years 2 and 3 than in year 1. In year 2 there was a reduced emphasis on scattering theory and perturbation theory, in keeping with the choice of textbook. Thus years 1 and 3 showed a higher score on scattering theory, as covered in questions 20--24. For instance, the $S$- and $T$-matrices were not explicitly covered in year 2. In year 3 we reinstated this material, and thus year 3 showed the highest score overall on the GQMCS.

\subsection{Textbook}
\label{ssec:textbook}

The second question we set out to address is, what graduate-level quantum mechanics textbooks, if any, incorporate all four periods of quantum mechanics in a pedagogically sound and readable way? To this end we reviewed over fifty graduate level quantum mechanics textbooks. We divided these textbooks into three categories. First, there are classic texts with which most people are familiar and have been used over many years, such as Landau and Lifshitz,\cite{Landau1977a} Schiff,\cite{schiff1968} and Sakurai.\cite{sakurai1994} Second, there are more specialized and applied texts which are well known to people in particular sub-fields of physics and sometimes used in present graduate courses, such as Peres,\cite{peres1995} Cohen and Tannoudji,\cite{Cohen-Tannoudji1977a} and Levi,\cite{leviAFJ2006} which are strongly directed toward quantum foundations, atomic, molecular, and optical physics, and engineering and chemical physics, respectively. Some of these are new, but did not satisfy the criterion of covering all four periods of quantum mechanics, and are not satisfactory for a general first year graduate quantum mechanics course. Third, there are new texts that cover all four periods of quantum mechanics, examples of which include Basdevant and Dalibard,\cite{basdevantJL2005} Rae,\cite{raeA2002} and Gottfried and Yan.\cite{gottfriedK2004} These texts have necessarily been written over the last twenty-five years, and mostly in the last five or ten years.

The newer textbooks take significantly different approaches from the classics. For instance, some incorporate the density matrix formalism and entanglement concepts from the very beginning. Most explicitly state the postulates of quantum mechanics. The question is which texts are most useful for implementing reforms as evidenced by explicit assessment tools, rather than individual or popular opinion.

In the three years of our study we used three different texts. In the first year, which we taught as a benchmark course, we used Sakurai,\cite{sakurai1994} together with a standard syllabus. In the second and third years we used two books from the third category, texts that cover all four periods of quantum mechanics: an English translation of the French text by Le Bellac,\cite{lebellac2006} and the second edition of Gottfried and Yan.\cite{gottfriedK2004} We chose Le Bellac as an example of a European text. The French education system does incorporate all four periods of quantum mechanics. However, most French texts are at too high a mathematical level to be useful to first year graduate students in the U.S.; there is a mismatch in the pace of mathematical education, which begins in elementary school and propagates forward all the way to graduate school. Le Bellac\cite{lebellac2006} is the lowest level French mathematical text of high quality that we could find and has many physical examples. We chose Gottfried and Yan\cite{gottfriedK2004} as an American text that covers all four periods of quantum mechanics and is both rigorous and complete within the constraints of what a typical U.S.\ graduate student can learn at this level.

In the following we evaluate the strengths and weaknesses of each text. The main source of our information is the assessment tools discussed in Sec.~\ref{sec:assessment}. In particular, the written student evaluations and the GQMCS provided the majority of the data.

\subsubsection{Benchmark Course: Sakurai}

One of the primary strengths of Sakurai is that the first two chapters are very readable, as students stated in interviews and written evaluations. The full immersion strategy of the text, which begins with the Stern-Gerlach experiment,\cite{gerlachW1922} followed by an inductive presentation of the basic elements of matrix formalism, ket notation, quantum dynamics, quantum-classical connections, and path integration, was appreciated by students in interviews near the start of the course. Other strengths of Sakurai are that it contains numerous physical examples in most sections, as students stated in interviews, and that approximation methods are couched in terms of applications, as students stated in written evaluations. A strength commented on only indirectly by students is that Sakurai does not begin with a historical review of quantum mechanics. That is, students commented negatively in student interviews and written evaluations on historical reviews in other texts. Moreover, students did very poorly on an initial assignment given in the third year based on Gottfried and Yan's historical review. We believe that this is because historical reviews assume a knowledge of basic concepts of undergraduate quantum mechanics, which research has shown that many students do not have.\cite{Steinberg1996a,Wittmann2002a,Wittmann2005a,McKagan2006a,Johnston1998a,Singh2001a,Sadaghiani2006a,Brookes2006a,Singh2008a}

Sakurai has significant weaknesses, mostly because it is dated. It covers mainly periods I and II of quantum mechanics. For instance, its presentation of the density matrix formalism suggests that density matrices are reserved for finite temperature quantum mechanics and must be related to thermal distributions. Most students in the first year were unable to give a meaningful answer to questions 7 and 10 of the GQMCS and corresponding questions on the final exam, which required a deeper understanding of density matrices to answer correctly. The inability to understand density matrices is a serious problem, which can lead to a lack of understanding of entanglement. Although Bell's inequalities are mentioned in Sakurai, comments in office hours and in class indicated that students considered them as a curious side-effect and, due to their lack of understanding of density matrices, were not able to understand technological applications of quantum mechanics to information processing, cryptography, etc. After using Sakurai, students did not conceive of the possibility of a set of well-defined postulates from which quantum mechanics can be derived, as was glaringly apparent on their Graduate Quantum Mechanics Conceptual Surveys in the answers to question 1. For instance, students in the first year wrote ``What postulates?'' in answer to this question, or ``Schrodinger equation?''.

Sakurai also has certain imbalances. For instance, one third of the text is on angular momentum. Is it the best use of time to teach students how to calculate arbitrary Clebsch-Gordan coefficients in a first year graduate quantum mechanics course? In Sec.~\ref{sec:content} we described how we removed part of this material to make room for a discussion of topics such as the postulates of quantum mechanics. A more serious imbalance is Sakurai's treatment of symmetry, which is very brief and not profound. This lack can be rectified in graduate group theory and quantum field theory courses, but students might not have either of these courses, because course requirements depend on the particular graduate program.

\subsubsection{First Reform Course: Le Bellac}

Because readers may be less familiar with new texts, we briefly outline Le Bellac. This text begins with a historical perspective. It then uses the strategy of first presenting quantum mechanics, from statics to dynamics to entanglement, solely in the context of the two state system, with an occasional segue to slightly larger finite dimensional systems. This presentation comprises the first six chapters and one third of the text. The text proceeds with a more standard treatment of infinite dimensional systems, including scattering theory, although without the path integral formalism. Finally, it has a set of applications to atomic, molecular, and optical physics, followed by a chapter on the dynamics of open quantum systems. We chose Le Bellac because it is a ``generalist'' text which tries to cover all areas of quantum mechanics accessible at this level. At the same time, it has the interesting strategy of staying with the two-state system significantly longer than other texts. We also wanted to try a European text.

Students gave mixed reviews to Le Bellac. In written evaluations they stated that it ``has poor explanations and homework problems that are more mathematical than physical'' and that it was ``extremely confusing.'' There were no overall positive comments on the textbook in written evaluations. In contrast, in interviews students were very enthusiastic about the many physical applications in Le Bellac, especially Chapter five, which treats quantum chemistry and other areas where finite dimensional systems can be analyzed in considerable detail. In these same interviews and as evidenced by the GQMCS, they responded well to the clear statement of the postulates of quantum mechanics, retaining this knowledge even a semester later.

For instance, a correct response to question 1 of the the GQMCS was ``In Ket notation: I. States are represented by a vector $|\psi\rangle$ in Hilbert space; II. Measurables are obtained through Hermitian operators; III. Time evolution is governed by the Schr\"odinger equation $\hat{H}|\psi\rangle = i \hbar \frac{\partial}{\partial t} |\psi\rangle$; IV. Probability of obtaining measurable $a$ from operator $\hat{A}$ is $|\langle \psi | \phi_a \rangle |^2$ where $\hat{A}|\phi_a \rangle = a |\phi_a\rangle$.'' A briefer correct response was simply ``States, Measurables, Probabilities, Time Evolution, Symmetrization.'' (We did not take off credit for postulates of single rather than many-body quantum mechanics, because the survey question is ambiguous.) Students were able to give detailed and correct explanations of entanglement concepts and open quantum systems on homework, exams, an independent project, and on survey questions 7 and 10. The chapter on entanglement in Le Bellac is seamlessly integrated with the treatment of two-state systems, not something we encountered in any other texts. In casual conversations with students one year later, several students, not the top in the class, appeared comfortable with these concepts and praised the text.

Even though we chose the least mathematical of the French texts, there was an educational mismatch between European and U.S.\ systems. Students almost uniformly complained in written evaluations about the level of mathematics, for instance, the treatment of unbounded operators. Le Bellac uses a spiral teaching method in which key topics are presented three times, each time in greater depth and with more rigor. The first time such topics are typically presented as statements of fact, without proof. Although this method seemed like a good idea when we selected the text, the students responded very badly to it. Some were even outright angry, as they expressed in office hours and on written evaluations. When asked for more details during a discussion in lecture, several stated they felt that a subject should be proved carefully from the beginning or they could not understand it. Physics graduate students really do want rigor and proofs.

Students found the presentation of two-state systems simultaneously in terms of polarization states of photons and spin-states of atoms (Stern-Gerlach) to be confusing, as they stated in interviews and was apparent in problem sets. Specifically, students were able to
answer questions about spin 1/2 systems, but were not able to answer equivalent
questions about photons, and they expressed confusion about the
geometry of photon polarization, were not able to answer questions about
photon polarization geometry, or made errors in geometry that led to
misunderstanding about polarization. Our interpretation of these problems is that students did not have a good grasp of polarization on which to build their understanding. Polarization states and spin 1/2 involve different geometries in their visualization; a boson which is a two-state system confuses students who believe that bosons are integer spin and therefore have an odd number of states. Also photons have a classical limit while spin 1/2 particles do not, a fact that students cannot understand at this level.

We were forced to supplement this text to satisfy the research needs of our students, not all of whom go on to take a quantum field theory course. We had to independently cover path integration. We also had to give a more thorough treatment of perturbation theory and scattering theory than the one in the text. In doing so we tried to follow as closely as possible the direction and intent of Le Bellac.

\subsubsection{Second Reform Course: Gottfried and Yan}

Although the first edition of Gottfried and Yan\cite{gottfriedK2004} is a classic text that covers mainly periods I and II, their second edition has a modern treatment that incorporates all four periods of quantum mechanics, integrating them throughout. Because readers might be unfamiliar with the second edition, we give a very brief overview. After giving a historical overview up through the present day, all the basic tools of quantum mechanics are presented with almost no physical examples. This presentation comprises one quarter of the text. The authors move on to a series of examples in the context of low-dimensional systems. They then discuss statistics, symmetries, and scattering theory for approximately half the text. The last quarter of the text contains advanced topics, including quantum fluctuations, a significant chapter on the interpretation of quantum mechanics, and an elementary treatment of the Dirac equation.

The breadth and rigor of the material in this text is far beyond those of most quantum mechanics texts we reviewed, including the other two we used in courses. Students did not complain about revisiting topics, as with Le Bellac, because topics were always presented in a rigorous, quantitative way. Students developed an understanding of entanglement and the postulates of quantum mechanics right from the beginning, as evidenced in problem set performance, a written project on the latest research, and student interviews, and were able to retain that understanding many months later, as evidenced by the GQMCS on questions 1, 7, and 10, and re-use of these concepts on the final exam in the second semester. In the second semester we were able to incorporate more advanced topics on interpretation of quantum mechanics and the Dirac equation as final projects, in which students studied the material on their own, including original scientific papers, wrote a report, and conceived and solved their own problem set on the topic. The idea of a final report in which students did something more closely akin to research was developed through the three years; it was only in the third year that we had the idea to make this report as extensive as three weeks of homework. Two students thought this was too much work; otherwise the response was very positive, with unsolicited positive remarks on the written evaluations.

Students gave mixed reviews to Gottfried and Yan. In evaluations they wrote that it was ``not very readable,'' ``very abstract for new concepts,'' and ``weak in their examples.'' Their principal complaint on written evaluations and in interviews was the lack of physical examples. Only one half to two thirds of the material in the text can possibly be covered in two semesters. Thus the chapters on foundations and basic tools ended up taking most of the first semester. We were forced to supplement this material with examples in order for the students not to be lost (as they informed us during student interviews in the third week of the course). The historical chapter is in fact quite advanced; we believe it might be more successful if it was assigned toward the end of the course, instead of at the beginning, as we did. Students did not find an initial foray into this material useful, as mentioned in our discussion of the use of Sakurai in the first year; an initial problem set in which students were supposed to summarize what they had learned in this chapter was a disaster, with only one or two students in the whole class able to write anything comprehensible. Finally, we were concerned that treatments of certain topics, for instance degenerate perturbation theory, were so general and abstract that they were difficult to relate to real examples. Such treatments are technically more complete but not necessarily more illuminating, as the students informed us both by comments and via exam performance.

\section{Teaching Methods}
\label{sec:methods}

PER has demonstrated that traditional lecture methods, in which the instructor stands at the blackboard and speaks for fifty minutes without pause while the students listen and take notes, are not particularly effective for student learning, and that interactive engagement methods, in which students participate more actively in class, lead to much greater student learning.\cite{Hake1998a,Hrepic2007a} Because the ineffectiveness of the traditional method is so well-established, we did not test this method on our students, even in the benchmark course. To make the course more interactive we incorporated the modifications discussed in the following in our teaching methods. Although we did not introduce specific systematic differences in teaching method from one year to the next, the extent of the interactive teaching methods gradually increased as we received more feedback from various forms of assessments. None of the methods used required any additional time to implement, an important feature for busy faculty.

First, the instructor elicited responses from every student in every lecture. This is reasonable only for small classes, but even in larger classes the instructor can interact with many students. Considering that we did so with twenty-two students in fifty minutes, this is a student-to-instructor interaction every couple of minutes during lecture. This interaction keeps students alert and involved. To encourage the students to participate in this way, in-class participation was part of the grade. Additional encouragement was provided by the following rules: questions early on in lecture were allowed to be answered by volunteers; when there were no more volunteers, we chose the remaining students arbitrarily for particular questions. Because students prefer to answer questions about which they have some idea, there was a strong motivation to volunteer at the earliest opportunity. A student was not allowed to answer a second question until all other students had answered.

Second, the instructor used several methods to encourage students to speak to one another during lecture. Students were asked to explain the answers to instructors' questions, which were posed in lecture notes ahead of time. The instructor sometimes sat with the students during lecture to generate a more informal atmosphere for discussion. Students were also asked to predict certain aspects of the lecture, for instance, to predict the course of a proof before going through it, or predict how one might implement important concepts mathematically. These discussions were often student to student. Finally, team homeworks were assigned for certain difficult numerical problems, in order to increase their \emph{esprit de corps}; this \emph{esprit de corps} spilled over into lecture. Grading was not on a curve, because the classes were too small to justify such a method. Thus students were not pushed to being competitive with one another.

Third, students were asked to break up into groups of two to four to work on problems. The classroom had boards on all four sides, so with the small class size we were able to ask students to write their results up on the boards. PER has demonstrated that students can learn more effectively through group work than working individually.\cite{Heller1992a} Teams were chosen by the students, but if teams became fixed over many lectures, we assigned new teams. No one-person teams were allowed. Early on the instructor actually left the room to help them not feel pressured. Later we discovered that simply sitting down on the side (not in front) was sufficient to remove any intimidation factor.

Fourth, there was a focus on concepts both in class and on exams. Every exam included 25\% conceptual questions to encourage students to think more deeply about quantum mechanics, rather than falling into the trap of a formulaic approach so endemic to earlier stages of physics education. We did not include conceptual questions in our homeworks, which we retrospectively believe to be an error. PER has demonstrated that although it is possible (and common) for students to do well on calculational problems but poorly on conceptual problems, the reverse is not true.\cite{Mazur1997a} Thus, conceptual understanding does not follow automatically from traditional problem-solving, but must be taught explicitly. An emphasis on conceptual understanding can even improve traditional problem-solving skills.\cite{Mazur1997a}

Fifth, all notes were provided online in advance. From one third to one half of the students read the notes ahead of time. We put open questions directly in the notes to guide their thinking. These open questions were a primary source of questions posed in class. Thus the students had a chance to think about some in-class questions in advance.

Sixth, long pauses were interspersed with the lecture, thereby giving the students time to think. For every ten minutes of lecture there were about two minutes of pauses. Research shows that slowing down the pace in this manner leads to improved student achievement on complex written tasks and student responses in class that are more frequent, longer, more complex, more likely to be supported by logical argument, and given by more students.\cite{Rowe1986a} This, together with sitting down and having open discussions, led to an engaged, informal lecture atmosphere.

Seventh, the instructor emphasized from the beginning of the course that he wished students to correct any error made on the board or in the presentation, no matter how small. Any student who did so was lauded. We believe this was a better model of both teaching and scientific inquiry than the air of absolute authority common to many lectures. Moreover, it kept the instructor on his toes.

Overall, the lecture style was Socratic in flavor. Students had a considerable degree of control over course content compared to more traditional lecture methods. Student response to these changes was very positive. Student attendance in lecture was 100\% except in case of illness over all three years. We increased use of these teaching innovations through the three years of our study. For instance, in the first year we did not especially encourage student-to-student interaction during lecture. We emphasize that none of these modifications ultimately took time away from the amount of material we were able to present.

\section{Assessment Tools}
\label{sec:assessment}

Most graduate-level instructors are research-focused tenured or tenure-track professors. Thus it is vital to choose assessment tools that do not take extra time beyond usual lecture preparation, grading, and office hours. The following seven assessments did not take any extra time. We found them highly useful for making incremental improvements in the course. In the following we discuss each of the assessment tools, which were cited as sources for the statements of Sec.~\ref{sec:content} and \ref{sec:methods}.

\subsection{Conceptual Surveys}

Conceptual understanding is a key part of becoming a professional physicist. Midterms and finals typically test calculation understanding, with conceptual understanding at best a peripheral issue. In addition to having an explicit conceptual component of our exams, we employed two assessment tools aimed specifically at conceptual understanding.

Before beginning our study, we supposed that physics instruction is a cumulative effort, especially in courses with the same title. Thus we thought that graduate quantum mechanics should build on and improve understanding of undergraduate quantum mechanics concepts. To test this hypothesis we gave the Quantum Mechanics Conceptual Survey (QMCS),\cite{QMCS,McKagan2008e} which is designed to test students' conceptual understanding of the basic ideas of quantum mechanics introduced in a sophomore-level modern physics course, as a pre-test and a post-test. In the first year of the course the post-test was given at the end of the first semester, and in subsequent years it was given at the end of the year.

The average score on the 12 common questions of the QMCS that were asked in all three years was $82\% \pm 3\%$ both pre and post. (All uncertainties are standard errors on the mean.) The average post-test scores for years 1--3 were $78\% \pm 6\%$, $91\% \pm 2\%$, and $76\% \pm 6\%$, respectively. None of these differences are statistically significant. Because they appear in both the pre-test and the post-test scores, any variations between the years are likely due to variations in the student populations, rather than due to instruction.

\begin{figure}[h!]
\includegraphics[width=\columnwidth]{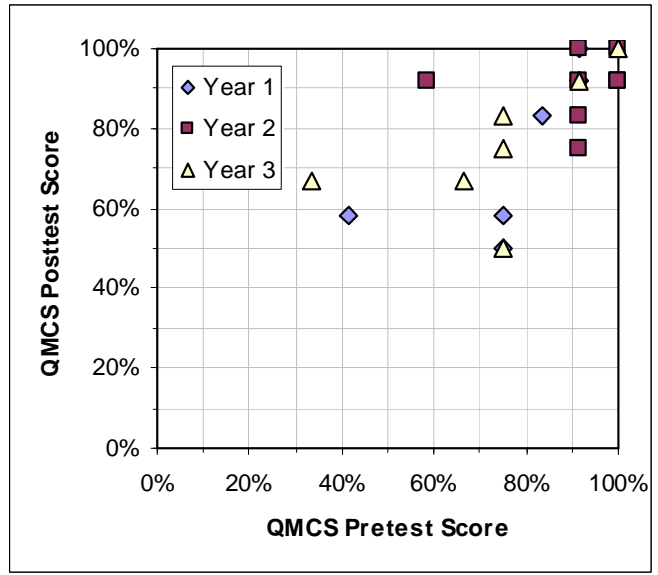}
\caption{The post-test versus pre-test scores on the QMCS show that taking a course in graduate quantum mechanics does not lead to significant gains in undergraduate conceptual understanding.} \label{QMCSprepost}
\end{figure}

The average gain was not statistically different from zero for any of the years. As can be seen in Fig.~\ref{QMCSprepost}, negative, positive, and zero gains were about equally common among individual students, and most students ($70\%$) had one point or less difference between the pre-test and the post-test. Although one might hope that students would continue to improve their basic conceptual understanding as a result of more advanced instruction, this result is not surprising because the concepts tested by the QMCS are not explicitly covered in graduate quantum mechanics. This lack of improvement in lower-level understanding in a higher-level course is consistent with earlier results showing that scores on the BEMA, a conceptual test of basic electricity and magnetism, do not improve after a junior-level E\&M course that focuses on advanced techniques.\cite{Pollock2007b}

\begin{figure}[h!]
\includegraphics[width=\columnwidth]{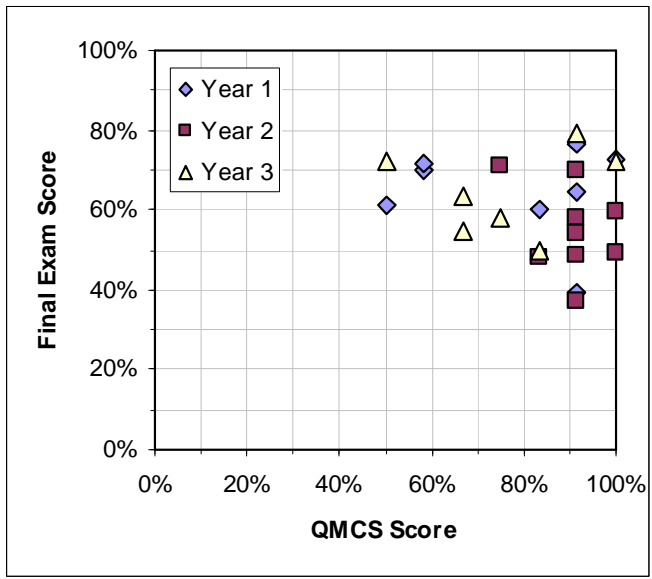}
\caption{Scores on final exam versus post-test scores on the QMCS. The lack of correlation shows that final exam performance in a graduate course is not indicative of undergraduate conceptual understanding. The Pearson correlation coefficient is $-0.21$ for the final exam versus QMCS post-test and $-0.0056$ for final exam versus QMCS pre-test.} \label{QMCS}
\end{figure}

Furthermore, Fig.~\ref{QMCS} shows that there is no significant correlation between final exam scores and QMCS scores, indicating that the QMCS is not a good predictor of success in graduate quantum mechanics.

In contrast, there do appear to be differences between QMCS scores for students with different backgrounds, for example, M.S.\ and Ph.D.\ students. For students who completed the entire course and took both the pre-test and the post-test, the average post-test score was $86\% \pm 5\%$ for M.S.\ students and $80\% \pm 5\%$ for Ph.D.\ students, compared with $88\% \pm 4\%$ for students enrolled in graduate quantum mechanics at the University of Colorado, and $55\% \pm 1\%$ for students after completing a traditionally taught modern physics course for physics majors at the University of Colorado. The average score for all three groups of students enrolled in graduate quantum mechanics is statistically different from the average score for the modern physics students (according to a two-tailed $t$-test, $p<0.001$). Although none of the other differences are statistically significant due to the small numbers of students involved, they are suggestive. The difference between the M.S.\ and Ph.D.\ students' scores is also expected, given that M.S.\ students were typically CSM undergraduates, so they came from a highly selective and rigorous program, whereas the Ph.D.\ students came from a variety of undergraduate institutions.

To assess student understanding of the material covered in graduate quantum mechanics we designed a new Graduate Quantum Mechanics Conceptual Survey (GQMCS), that we used as a post-test of material covered in the course. This survey (see Appendix~B) is essay style, and consists of twenty-four questions. It was given as an extra-credit pre-final at the end of the second semester. Partial credit was allowed. We were very concerned about a bias being introduced by students also taking the exam at the beginning of the course. We were also concerned about potential bias in our own lectures. Therefore, after writing the exam before the first year, we did not look at it again until after the students had taken it. Students were given 2 hours outside of class to complete the survey, so they had an average of five minutes to answer each question.

The average scores on the GQMCS for years 1--3 were $68\% \pm 5\%$, $64\% \pm 3\%$, and $78\% \pm 5\%$, respectively. The difference between years 2 and 3 is weakly statistically significant, according to a two-tailed t-test ($p=0.05$). We did not find that the average scores were particularly useful to compare student understanding, as different courses led to different proficiency on different questions. Instead, we compared answers to individual questions to provide the underpinnings of our analysis in Sec.~\ref{sec:content}. However, we note that the third year, which provided the most comprehensive course with the most comprehensive textbook, did lead to the highest average, despite student complaints about the text.

The results of the GQMCS closely followed textbook choice and course content. For instance, students taking the Sakurai-based course had no idea what the postulates of quantum mechanics are. They even wrote things like ``What postulates?'' or ``There are no postulates of quantum mechanics.'' This tool was mainly useful in validating our hypothesis that students did not acquire any significant understanding of periods III and IV in the benchmark course, and in determining where the reformed courses in years two and three were lacking. For instance, students using Le Bellac did not have as good an understanding of scattering theory as those using Sakurai or Gottfried and Yan, as can be predicted by examining these textbooks.

One conclusion suggested by these surveys is that students did not take away any information that was presented solely in the textbook. Assigned reading, even with the occasional homework problem to help encourage students to actually do it led to a score of zero on the relevant problem on the survey when not supplemented with a lecture on the topic. This results strongly suggests the need to cover periods~III and IV in lecture, rather than try to include them via study outside of the classroom. At the same time, this result highlights how useful lectures can be when they are interactive.

Another point is that concepts, unlike calculational knowledge, were often all or nothing. Students mostly either understood the concepts well or did not really understand them at all. Despite offering partial credit, it was rare that partial credit was given for a reason other than the answer being too brief. For example, in the first year students could not list any postulates; in the second and third years all students listed all the postulates. In the third year all but one student gave a blank or a completely wrong answer to question~8 concerning the connection between SU(2) and SO(3), in contrast to years one and two, when most students got full credit. (The difference was likely due to the lecturer not spending as much time on the subject as in the first and second years.) In the third year students either retained total knowledge of Clebsch-Gordan coefficients or almost none.

\begin{figure}[t]
\includegraphics[width=\columnwidth]{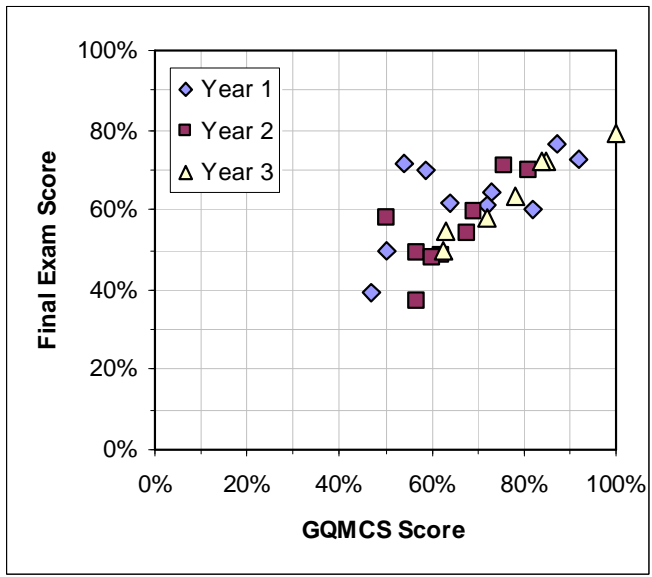}
\caption{The strong correlation between the scores on the final exam and the scores on the GQMCS show that conceptual knowledge of graduate topics is an excellent indicator of calculational ability. The Pearson correlation coefficient is $0.74$.} \label{GQMCS}
\end{figure}

As shown in Fig.~\ref{GQMCS}, performance on the GQMCS was a very good indicator of performance on the final, with a Pearson correlation coefficient of $0.74$, despite the fact that the final was 75\% calculations. This result reinforces other research demonstrating that conceptual knowledge can improve the student's ability to perform calculations.\cite{Mazur1997a}

\subsection{Student evaluations}

We used three kinds of student evaluations. First, a university-wide assessment tool is required at our university. Students give a ``grade'' to the professor and answer specific questions, mainly about the mechanics of the class: how well it was graded, whether or not the course objectives were clear, the availability of the professor, and whether or not the professor knew the material. There are 14 core questions and 3 department-specific questions. Our grade was at or above the department average in most categories. Although this assessment tool was useful for evaluation of the instructor's ability to run a course, it was not particularly useful for evaluating the effectiveness of reforms, because the questions asked were too general, for instance, ``course material is well presented,'' rated on a five-point scale. We have included this assessment tool for completeness.

Second, we provided a written essay-style evaluation. We found this evaluation to be much more useful because it allowed students to be specific. The instructor took these evaluations seriously and modified the course from semester to semester according to students' recommendations. The instructor informed students ahead of time of this fact. The six questions were as follows. (1) What were three strong points of the course? (2) What were three weak points of the course? (3) How could lectures be improved? (4) How could the homework be improved? (5) How could the exams be improved? (6) Any other overall comments?

Retrospectively, we ought to have also asked about the textbook. However, most students commented on the textbook anyway in questions (1), (2), and (6), although only in the negative. In general, students wrote long and detailed responses. The material learned from these evaluations was summarized in Sec.~\ref{sec:content} and Sec.~\ref{sec:methods}. We briefly summarize a few points again here. Students had a very high level of appreciation for improved teaching techniques. Students consistently wrote how much they enjoyed the lecturing style and how it kept them engaged. Each novel teaching technique was noticed by the students and commented on in the evaluations: pacing, student-to-student and student-to-instructor interactions, etc. They also frequently asked for more physical examples over all three years. Students often claimed that without applications they had not learned anything.

Our third student evaluation was performed as follows. In the third and fourth week of the course the instructor required all students to come to office hours. At CSM five office hours per week are required. However, two or three office hours would be sufficient for many graduate courses. We note that student expectations of the instructor at CSM are very high; even five office hours is frequently commented on as being ``unavailable.'' The instructor spent ten minutes with each student. In particular, students were invited to state what they wanted from the course and to comment on how the course was progressing. Then the instructor listened. Interviews were concluded by inviting the student to give feedback at any time, either anonymously or directly, and thanking her/him for her/his comments.

The interviews allowed for a mid-course correction, rather than having to wait for final evaluations and then making corrections in the following semester or following year. For instance, when students almost unanimously stated that Gottfried and Yan did not have enough physical examples, we were able to add these to our lectures and homeworks immediately. In contrast to written student evaluations, student interviews tended to produce more positive comments about textbooks. This difference is interesting, especially because students were not explicitly asked to talk about the textbook in either the written evaluations or student interviews.

Another example of a mid-course correction was the addition of final projects on interesting new areas of physics in the third year of the course. This addition was based on both positive and negative feedback from the students. They complained about not having seen enough applications. At the same time they appreciated the solicitation of research objectives and course objectives on the first day of class by the instructor. Therefore, a lower-level writing assignment from the first and second years was adapted into a final research project.

Student interviews contained much more detailed information than written evaluations. Students tended to express more positive information, compared to written evaluations. For instance, they spoke positively about the textbook in interviews and negatively in written evaluations, as we have mentioned. These positive and negative comments provided complementary points of view. The interviews came near the beginning of the semester, and provide different snapshots in time; at the end of a course it is our observation that students are weary, especially in the second semester of the first year of graduate school. During interviews students believe they can still make a difference in a way that will affect them personally.

\subsection{Faculty Evaluations}

There were two kinds of faculty evaluations. The CSM physics department requires a yearly in-lecture visit from the department head to evaluate and improve teaching. Comments here included taking time to summarize the lecture at the end rather than rushing to finish a proof or other result and improving the instructor's handwriting, neither of which students commented on.

An alternate point of view was provided by a pair of physics education researchers visiting the lecture. Here the instructor learned to get students to explain their answers to other students during lecture, thereby increasing student-to-student interactions. This assessment was also extremely useful as a validation of teaching techniques the instructor implemented, as the PER evaluators had no prior knowledge of which techniques were being actively used. These evaluators also suggested that the instructor give students a question to consider for one to two minutes every fifteen minutes to allay exhaustion due to the intensive interactive nature of the instructor's lecture style. The implementation of this suggestion resulted in more open discussions even when not explicitly asked for by the instructor. We believe this increase is due to the students feeling both more accustomed to in-class discussion and less overwhelmed.

The main point is that faculty have a very different point of view from students, and one's own department is a considerable resource in improving a course.

\section{Discussion and Conclusions}
\label{sec:conclusions}

Because we changed both the content and the pedagogy of the course at the same time, it is difficult to isolate the effects of either change. However, the results are useful to inspire future, more controlled studies, and suggest immediate possibilities for course reform. Such reform is badly needed to help students become effective researchers in the field of quantum mechanics. To our knowledge the only other effort specific to first year graduate quantum mechanics was published by Singh just as we completed our own work,\cite{Singh2008a} and treats undergraduate quantum mechanics concepts. Singh's investigation deals primarily with the functional representation of quantum mechanics and visualization, as can be observed in her survey questions. Our survey, presented in Appendix~B, focuses on graduate skills, such as the Dyson series and path integration. Our study addressed all aspects of first year graduate quantum mechanics education, as they all need reform.

In addressing our first question concerning the inclusion of periods I--IV and associated course content change, there are several areas where significant further improvement can be made in course content. One difficulty with our present syllabus is that the low dimensional systems are presented primarily in terms of abstract, matrix, and ket notation, and incorporate entanglement concepts, while high dimensional systems are presented in terms of position representation and return the students to thinking in terms of wave functions. There is a disconnect between these two portions of the course, which needs further work.

The question of how to best represent the two-state system, whether as polarization states of a photon, as a spin 1/2 particle, as a two-state atom, or as a qubit, is an interesting research topic. The same can be said for how long one should constrain the discussion to the two-state system; the three textbooks we investigated all had different strategies in this regard. We have a personal preference for Le Bellac's strategy, as it allows one to present most aspects of first-year graduate-level quantum mechanics in a simple language that our more talented undergraduates could easily grasp. However, we recommend only one physical implementation of a two-state system, as Sakurai uses, that is, only spin 1/2, not polarization and spin 1/2 at the same time. Further research in this direction could be highly useful to creating an optimal course.

We found that it is possible to include periods III and IV of quantum mechanics without undue sacrifices. One professor in our department was concerned about the loss of the full version of the Wigner-Eckart theorem, but otherwise faculty were amenable to our changes. An important point about core course modification is it is better to have some overall faculty input. Then PER projects are less likely to be viewed as threatening and more as a department-wide effort. Approval was sought at various points through our three year study, beginning with the department head before the first year was taught.

In addressing our second question concerning the use of contemporary texts in Sec.~\ref{ssec:textbook}, there is some influence from the instructor, who presents the text according to his or her own style of teaching, which in turn affects the evaluations of the text. For instance, we gave a much stronger emphasis on concepts than other graduate quantum mechanics courses with which we are familiar. Therefore, it is preferable to repeat this study with different instructors. However, most instructors are not willing to write three sets of notes in three subsequent years with three different texts. Therefore, it is difficult to obtain this kind of data.

The main answer to our second question is that, although there is not yet a perfect text to achieve our goals, existing texts can be supplemented appropriately to do so. We hope that pedagogically improved texts that incorporate all four periods of quantum mechanics will appear in the near future. In the meantime, our assessment tools show that modern texts such as Le Bellac and Gottfried and Yan improve students' ability to develop important skills in quantum mechanics in an overall context, rather than as a grab-bag of methods. This improvement was noticeable in the responses on the GQMCS, in one-on-one conversations with students in the years following their taking their course, and in their performance on the conceptual portion of the final exam. Moreover, students feel more excited about the connection to modern research, and this excitement increases their overall interest in the course material, as they stated in written evaluations -- a strategy that can be applied to any course.

Our third question asked whether or not research-based teaching innovations from undergraduate courses could be reasonably and successfully employed in graduate courses. The answer is a resounding yes. Students were very responsive to Socratic and other modifications. Attendance was 100\% except in case of illness or attending a professional conference. Students formed a real \emph{esprit de corps}, and felt they had a say in their own education, according to their evaluations. This part of our study built continually over the three years. Because we were not willing to go back and teach in a traditional manner for the benchmark course, we cannot clearly demonstrate that students are better able to solve problems or understand concepts than otherwise. However, the result from the GQMCS that students showed an understanding of topics covered in lecture but not topics covered in the text alone suggests that the methods used in lecture were effective for student learning. Further, there is evidence that positive student attitudes are correlated with learning.\cite{Perkins2005a,Perkins2006b} A comparative analysis of the effect of different teaching techniques in graduate quantum mechanics remains a subject of future study.

Our fourth question addressed assessments of the first three questions. Besides midterms and finals, our assessment tools fell into three categories: conceptual surveys, student evaluations, and faculty evaluations. We found all three non-standard sets of assessment tools to be useful. Highly active research faculty are jealous of their time; an important criterion for assessment tools was that they should not take any extra time beyond office hours and usual lecture preparation. Therefore we did not employ clickers, ``Just in Time Teaching,'' or similar teaching and assessment methods, which require an initial extra effort comparable to developing a new set of lecture notes.

We found that students did not improve their understanding of quantum mechanics concepts at the modern physics level, as measured by the QMCS, while taking a graduate course. Course content and textbook made the largest difference in forming a new understanding of graduate quantum mechanics. Our teaching methods evolved considerably during the three years. Yet the results of this survey did not correlate with continual improvement of teaching methods.

Because our GQMCS was directed toward our own concept of what is necessary to do successful research, it can be argued that we ``taught to the test.'' However, as any instructor who has ever been disappointed in exam results knows, teaching to the test doesn't always work. Although it would be preferable to have a comprehensive assessment tool agreed upon by a larger number of researchers, our survey can serve as a first step toward that goal.

One goal of our study which we did not explicitly test is how well our course prepared graduate students for thesis research. An assessment tool to address this point could be developed. In particular, a follow-up study just after the students' thesis defenses, whether M.S.\ or Ph.D., would be very useful, either in the form of personal interviews or an online survey.

In conclusion, we have presented a number of ways in which graduate quantum mechanics can be reformed, in terms of course content, teaching methods, and assessment tools. None of these areas took significant extra time over and above the usual amount allotted to course preparation, and all led to improvements in student learning, attitude, and assessment.

\begin{acknowledgements}

We thank the Colorado School of Mines physics students for
participation in this study. We thank the University of
Colorado Physics Education Research Group, and the Colorado School of Mines
Physics Faculty for useful suggestions. We are grateful to Stephanie Chasteen,
Noah Finkelstein, Patrick Kohl, Vincent Kuo, and Carl Wieman
for helpful discussions and feedback.
This work was supported by the National Science
Foundation under Grant PHY-0547845 as part of the NSF CAREER program (LDC) and under Grant PHY-1542800 (SBM).
\end{acknowledgements}

\section*{Appendix A: Final Syllabus}
\label{app:syllabus}

\begin{enumerate}

\item Semester I

\subitem (1) Stern-Gerlach/qubits.
\subitem (2) Formal framework:
Hilbert space; matrix, ket, and functional representations; Mixed vs.\ pure states, entropy, entanglement, temperature; uncertainty principle; quantum-classical connections; postulates of quantum mechanics.
\subitem (3) Symmetries and conservation laws:
Translation, Parity, vector and tensor operators, rotation, spherical harmonics.
\subitem (4) Basic applications:
Benzene, NMR, Solid state theory.
\subitem (5) Angular momentum:
Orbital, Spin, Addition, Wigner-Eckart theorem (vector operators only).
\subitem (6) Basic approximation methods:
Perturbation theory, variational method, minimal examples.
\subitem (7) NB: Harmonic oscillator presented in year 2 only, could be re-introduced.

\item Semester II

\subitem (1) Propagators and path integration.
\subitem (2) Harmonic oscillator (year 3 only):
Creation/Destruction operators, coherent states, classical-quantum connections, equations of motion.
\subitem (3) Baby quantum field theory:
scalar 1D quantum field, quantum fluctuations of electromagnetic field, Casimir-Polder effect.
\subitem (4) Gauge transforms, Ahranov-Bohm Effect.
\subitem (5) WKB.
\subitem (6) Hydrogenic atoms:
Stark, Zeeman, Spin-orbit, van der Waals
\subitem (7) Identical particles, Helium atom.
\subitem (8) Advanced applications:
Photoelectric effect, resonance states, second deeper treatment of time-dependent perturbation theory
symmetry.
\subitem (9) Wigner's theorem, time-reversal, overview of all symmetries.
\subitem (10) Scattering theory:
Calculating cross sections, Lippman-Schwinger equation, Born approximation, partial waves, T and S matrices, time-dependent formulation via propagators, inelastic scattering.

\end{enumerate}

\section*{Appendix B: Graduate Quantum Mechanics Conceptual Survey}
\label{app:assessment}

The GQMCS was couched in the form of a ``practice final''
as follows.

\begin{center}
Physics 521 -- Graduate Quantum Mechanics II

Practice Final Exam

Time allowed: 120 minutes.

Total points: 100.

\end{center}

The primary purpose of this practice exam is to give you an overview
of what you should know at this level of quantum mechanics. The
secondary purpose is to help you study for your final. Although the
following is almost purely conceptual, the explicit conceptual
component of the final will be 20--25\%. However, the following
concepts \emph{underlie} everything you will calculate on the final.
Please note that material from the first semester of the course will
not be covered explicitly on the final, although, again it underlies
your calculations, and therefore appears in the following. Try to
answer all questions, even if the material was not explicitly
covered in class. Each question should take you five minutes or
less.

\begin{enumerate}

\item What are the fundamental postulates of quantum mechanics?

\item What is the meaning of the Schrodinger wavefunction?

\item What are the mathematical definition and physical meaning of hermitian and unitary operators, respectively?

\item What is a quantum measurement? Give an example of a sequence of two measurements on an initially degenerate state.

\item Give a physical example of a finite discrete, an infinite discrete, and an infinite continuous state space, respectively. What is the dimension of each space?

\item Write the general form of a translation operator. What are the essential mathematical properties of such an operator? What is a particular instance of such an operator? How can one determine if the Hamiltonian is symmetric under such a translation?

\item What is an entangled state? Give an example of two particle and three particle entangled states based on direct products. What is a cat state? Give an explicit example.

\item Explain the physical meaning of SU(2) and SO(3). How are they connected?

\item Explain the difference between a coherent state and an number state (here, number of energy quanta) for a harmonic oscillator.

\item Define the density matrix. What are its zero and infinite temperature limits? Define entropy in the density matrix formalism.

\item What is the mathematical definition and physical meaning of a path integral?

\item Explain the difference between the Schrodinger, Heisenberg, and Interaction pictures. When is each one used?

\item Write down the general form of the quantum mechanical rotation operator. What is the generator of rotation?

\item Explain the basic idea of Clebsch-Gordan coefficients.

\item Explain what time-reversal is. Are Maxwell's equations time-invariant? How about the Schrodinger equation? Justify your answer to the latter.

\item Compare and contrast the following approximation methods. Name at least one physical situation in which each method might be used.

(a) Semiclassical WKB

(b) Variational

(c) Non-degenerate time-independent perturbation theory

(d) Degenerate time-independent perturbation theory

(e) Time-dependent perturbation theory

\item Explain what the Dyson series is. Give an example of its application as a mathematical technique.

\item Explain the concept of resonance in exactly solvable two state systems.

\item Write down the Slater determinant and explain its meaning. What is its relation to the symmetrization postulate?

\item Explain the form of spherical Bessel functions of the first, second, and third kinds, both physically in terms of the scattering problem in three dimensions and mathematically as solutions to some version of the Schr\"odinger equation in three dimensions.

\item Define the cross section, the differential cross section, the scattering amplitude, and the scattering phase shift. How does one calculate cross sections?

\item Sketch the poles in the scattering amplitude for a generic potential that contains both bound and quasi-bound states. Explain what a quasi-bound state is and how one would estimate its lifetime.

\item Write down the Lippmann-Schwinger equation and indicate how it may be derived.

\item What is the S-matrix? The T-matrix? What are their physical meanings?

\end{enumerate}

\end{document}